\documentclass[prl,twocolumn,showpacs]{revtex4-1}

\usepackage{graphics,graphicx,amssymb,amsmath,color}
\begin{document}

\title{Polaron-mediated spin correlations in metallic and insulating La$_{1-x}A_{x}$MnO$_{3}$ ($A$=Ca, Sr, or Ba)}

\author{Joel~S.~Helton$^{1,2\ast}$}
\author{Daniel~M.~Pajerowski$^2$}
\author{Yiming~Qiu$^{2,3}$}
\author{Yang~Zhao$^{2,3}$}
\author{Dmitry~A.~Shulyatev$^4$}
\author{Yakov~M.~Mukovskii$^{4\dagger}$}
\author{Georgii~L.~Bychkov$^{5\dagger}$}
\author{Sergei~N.~Barilo$^5$}
\author{Jeffrey~W.~ Lynn$^{2\ddag}$}

\affiliation{$1$ Department of Physics, United States Naval Academy, Annapolis, Maryland 21402, USA}
\affiliation{$2$ NIST Center for Neutron Research, National Institute of Standards and Technology, Gaithersburg, Maryland 20899, USA}
\affiliation{$3$ Department of Materials Science and Engineering, University of Maryland, College Park, Maryland 20742, USA}
\affiliation{$4$ National University of Science and Technology ``MISiS," Moscow 119991, Russia}
\affiliation{$5$ Institute of Solid State and Semiconductor Physics, Academy of Sciences, Minsk 220072, Belarus}

\date{\today}
\begin{abstract}
Neutron spectroscopy measurements reveal short-range spin correlations near and above the ferromagnetic-paramagnetic phase transition in manganite materials of the form La$_{1-x}A_{x}$MnO$_{3}$, including samples with an insulating ground state as well as colossal magnetoresistive samples with a metallic ground state.  Quasielastic magnetic scattering is revealed that forms clear ridges running along the [100]-type directions in momentum space.  A simple model consisting of a conduction electron hopping between spin polarized Mn ions that becomes self-trapped after a few hops captures the essential physics of this magnetic component of the scattering.  We associate this scattering component with the magnetic part of diffuse polarons, as we observe a temperature dependence similar to that of the diffuse structural scattering arising from individual polarons.

\end{abstract}
\maketitle

\section{Introduction}
Complex oxide materials with the perovskite crystal structure exhibit a wide range of unique physical properties arising in part from intrinsic spatial inhomogeneity such as phase separation and nanoscale polarons in colossal magnetoresistive manganites \cite{Dagotto,Dagotto2005,Moreo}, polar nanoregions in relaxor ferroelectrics \cite{Gehring,Gehring2}, and spin and charge ordered stripes in cuprate superconductors \cite{Tranquada,Kivelson}.  In hole-doped manganites of the form La$_{1-x}A_x$MnO$_3$ (where $A$ is a divalent alkaline-earth ion such as Ca, Ba, or Sr) this intrinsic spatial inhomogeneity arises from the strong interplay of the spin, charge, lattice \cite{Millis1995,Millis1996}, and orbital \cite{Tokura} degrees of freedom.  For hole doping in the range of 0.22 to 0.50 the ground state is a magnetically isotropic ferromagnetic metal \cite{Okuda}, in which Zener double-exchange \cite{Zener} links the spin and charge degrees of freedom as $e_g$ conduction electrons hopping from one Mn ion to another provide both conductivity and ferromagnetic exchange (due to the strong Hund's rule coupling between the itinerant $e_g$ electrons and the localized $t_{2g}$ core electrons).  Colossal magnetoresistance (CMR) is observed at the combined ferromagnetic and metal-insulator phase transition, with $T_{c}\approx$265~K for optimally doped La$_{1-x}$Ca$_{x}$MnO$_3$ \cite{Schiffer,Cheong}.

Nanoscale polarons, consisting of a conduction electron and its associated local Jahn-Teller distortion, have been invoked to help explain many of the unusual properties of the manganite materials \cite{Jaime,Savosta,Lynn1996}, including their sensitivity to applied magnetic fields \cite{deTeresa,Souza,Moshnyaga,Booth}.  Polaron correlations have been observed at reduced wave vectors of $q=(\frac{1}{4}~\frac{1}{4}~0)$ and equivalent positions (with respect to the pseudocubic unit cell), associated with the CE-type charge and orbital ordering \cite{Goodenough} of half-doped La$_{0.5}$Ca$_{0.5}$MnO$_{3}$.  However, the observed polaron scattering is only structural in origin, while the expected magentic component of the polaron scattering has been elusive \cite{Adams2000}.  Near-optimally doped La$_{0.7}$Ca$_{0.3}$MnO$_{3}$ (LCMO) displays both static \cite{Adams2000,Dai2000} and dynamic \cite{Lynn2007} polaron correlations; it is the formation of static polaron order that truncates the ferromagnetic metallic phase in a weakly first-order phase transition \cite{Adams2004}.  La$_{0.7}$Sr$_{0.3}$MnO$_{3}$ (LSMO) and La$_{0.7}$Ba$_{0.3}$MnO$_{3}$ (LBMO), which display more conventional second-order phase transitions and a less pronounced CMR effect, feature only dynamic polaron correlations \cite{Chen}. Static polaron correlations are also observed near $T_c$ in La$_{0.8}$Ca$_{0.2}$MnO$_{3}$, which displays a ferromagnetic insulating ground state \cite{Dai2000}.  In neutron scattering measurements of $x$=0.3 LCMO \cite{Adams2000}, the intensity of the correlated polaron peak maximizes at a temperature near $T_c$ in a manner very similar to the bulk resistivity.  The intensity of diffuse scattering near a Bragg peak, associated with scattering from uncorrelated polarons, also increases rapidly with increasing temperature as $T_c$ is approached from below; however, this intensity features only a weak temperature dependence for higher temperatures.  This suggests that the number of polarons increases rapidly as the temperature approaches $T_c$, but as the temperature increases above $T_c$ the number of polarons remains roughly constant while the strength of correlations between these polarons decreases.
\begin{figure*}
\centering
\vspace{-45mm}
\includegraphics[width=16cm] {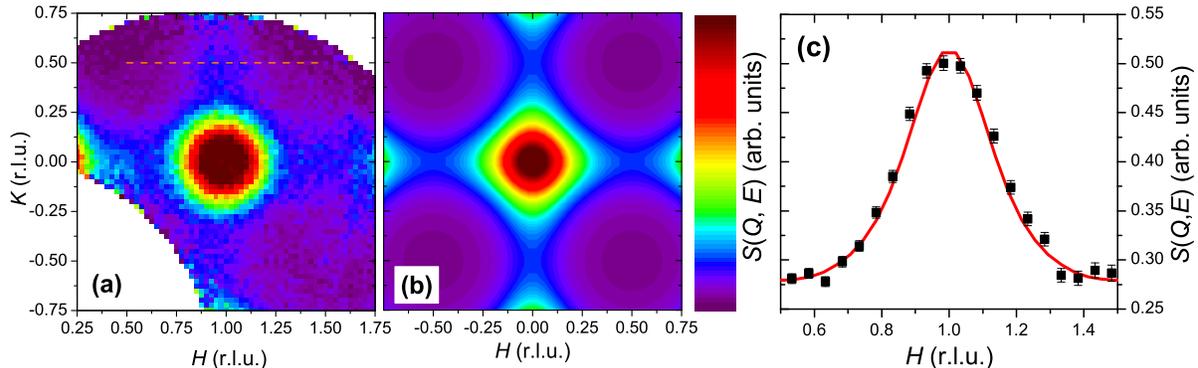} \vspace{-35mm}
\caption{(a) $S(Q, E)$ for La$_{0.8}$Ca$_{0.2}$MnO$_3$ measured at $T$=270~K using the DCS spectrometer.  The energy transfer has been integrated over $-3.0$~meV~$\leq~E~\leq~-$0.25~meV. The dashed line represents the scan direction for panel (c).  (b) Energy-integrated magnetic scattering calculated for ferromagnetic correlations based along hopping pathways as described in the text.  A correlation length of  $\xi$=0.7 hops provides a semiquantitative description of the experimental data.  The color scale is linear.  (c) ($H$~0.5~0) transverse scan through the center of a ridge, integrated over 0.3~$\leq~K~\leq$~0.7.  The red line is taken from the calculation.  The scale factor and background have been chosen to match the data.}
\label{Model}
\end{figure*}

A previous report \cite{Helton} on the dynamic spin correlations of near-optimally doped La$_{0.7}$Ca$_{0.3}$MnO$_{3}$ in the paramagnetic phase at a temperature of 265~K ($\approx$~1.03~$T_{C}$) found that two distinct components to the spin fluctuation spectrum could be observed in low-energy constant-$E$ neutron spectroscopy measurements.  A spin-diffusive central component \cite{Lynn1996,Dai2000} is observed as quasielastic scattering with a Lorentzian line shape where the energy half-width at half-maximum (HWHM) depends on the reduced wave vector as $\Gamma=\Lambda q^2$.  This field-dependent \cite{Lynn1997} component in samples near optimal doping displays a temperature dependence that closely matches the behavior both of the bulk resistivity and of static polaron correlations \cite{Adams2000}.  This spin-diffusive component is a unifying feature of La$_{1-x}$Ca$_{x}$MnO$_{3}$ materials as the spin-diffusion coefficient ($\Lambda$) increases smoothly with increasing doping, while the spin wave stiffness coefficient increases discontinuously by a factor of three when the doping crosses into the range of metallic compounds ($x\geq0.22$) \cite{Dai2001}.  In low-energy constant-$E$ measurements this spin-diffusive component appears as a broad ring of intense scattering surrounding the Bragg position. The second component in the paramagnetic scattering takes the form of ridges of strong quasielastic scattering intensity connecting the rings of diffusive scattering along ($H$~0~0) and equivalent directions.  Given the coupling between the magnetic and lattice degrees of freedom in these materials, it is possible that this component is partly structural in origin.  However, this component must be primarily magnetic given that the ridge intensity decreases with increasing $Q$ proportional to the square of the Mn form factor.  We will show that these ridges of scattering represent a unique facet of dynamic short-range spin correlation that is universal among ferromagnetic La$_{1-x}A_{x}$MnO$_{3}$ materials with both metallic and insulating ground states and that represents the magnetic component of the nanoscale polarons.

\section{Results}
Broad ridges of magnetic scattering running along the [100]-type directions are indicative of short-range ferromagnetic correlations along directions parallel to the $a$, $b$, and $c$ principal axes (the Mn nearest-neighbor directions).  The ferromagnetic exchange in these materials arises from Zener double exchange, related to the hopping of itinerant $e_g$ electrons between Mn ions.  If the carrier hopping pathways are restricted to nearest neighbors only, then short-range ferromagnetic correlations could result where the strength of the correlation depends on the number of hops connecting two Mn ions rather than the distance between them.  This leads to preferential ferromagnetic correlation along the principal axes.  Figure~\ref{Model}(a) shows $S(Q, E)$ for $x$=0.2 LCMO at 270~K in the ($H$~$K$~0) scattering plane surrounding the (1~0~0) Bragg position.  The energy transfer has been integrated over $-3.0$~meV~$\leq~E~\leq~-$0.25~meV in order to approximate the energy-integrated scattering intensity; the neutron energy gain data have been chosen due to lower background, as detailed balance plays little role at this temperature and energy transfer range.  Figure~\ref{Model}(b) displays the calculated magnetic scattering intensity for short-range ferromagnetic correlations dependent on the number of nearest-neighbor hops connecting two Mn ions; the modeled correlation decays exponentially as $\langle S_i S_j\rangle\propto e^{-n/\xi}$, where $n$ is the number of hops connecting ions $i$ and $j$ and $\xi$ is a correlation length. Figure~\ref{Model}(c) displays the data for a transverse scan through the ridge centered at (1~0.5~0) along with the calculated magnetic scattering intensity for this scan.  The similarity between the data and this simple model suggests that the magnetic scattering ridges observed in these materials can be thought of as the magnetic part of the diffuse polaron scattering, arising from the short-range ferromagnetic correlations created when a conduction electron and its associated (polaron) lattice distortion become self-trapped after only a few nearest-neighbor hops between spin-aligned Mn ions.

We have performed neutron measurements on six single crystal samples of La$_{1-x}A_{x}$MnO$_{3}$ where the divalent alkaline-earth metal ion $A$~=~Ca, Sr, or Ba and 0.15~$\leq$~$x$~$\leq$~0.30 spans from materials with a ferromagnetic insulting ground state ($x$~$\leq$~0.22) to colossal magnetoresistive materials with a ferromagnetic metallic ground state (optimal doping at $x$~$\approx$~1/3).  The measured La$_{1-x}$Ca$_{x}$MnO$_{3}$ samples were $x$~=~0.15 ($m$=1.4~g, $T_c$=160~K), $x$~=~0.2 ($m$=0.7~g, $T_c$=181~K) \cite{Adams2004}, $x$~=~0.25 ($m$=0.7~g, $T_c$=187~K), and $x$~=~0.3 ($m$=1.5~g, $T_c$=257~K) \cite{Adams2000,Lynn2001}.  Near-optimally doped samples of La$_{0.7}$Ba$_{0.3}$MnO$_{3}$ ($m$~=1~g, $T_{c}$~=~336~K) \cite{Barilo} and La$_{0.7}$Sr$_{0.3}$MnO$_{3}$ ($m$~=~3~g, $T_{c}$~=~351~K) \cite{VasiliuDoloc1998} were also measured. Inelastic neutron spectroscopy studies on all six samples were carried out using the BT-7 thermal triple-axis spectrometer at the NIST Center for Neutron Research (NCNR) \cite{BT7}.  Additionally, high-resolution [energy full width at half-maximum (FWHM) of $\approx$230~$\mu$eV] time-of-flight measurements were performed on $x$=0.2 and $x$=0.3 LCMO using the Disk Chopper Spectrometer (DCS) at the NCNR.  All samples were mounted with the ($H$~$K$~0) scattering plane horizontal. The samples are indexed with the pseudo cubic unit cell where the lattice spacing, equivalent to the nearest-neighbor distance between Mn ions, is approximately 3.9~{\AA}. Neutron-scattering measurements using the BT-7 thermal triple-axis spectrometer were performed with a fixed neutron final energy of 14.7~meV.  Pyrolytic graphite filters were placed in the scattered beam to reduce contamination from higher-order neutron wavelengths.  The spectrometer was configured with an 80$^{\prime}$ radial collimator between the sample and the horizontally focused analyzer and either a 50$^{\prime}$ (for $x$=0.3~LCMO) or 80$^{\prime}$ (all other samples) Soller collimator between the monochromator and the sample.  Time-of-flight measurements using the DCS spectrometer were performed with an incident neutron wavelength of 3.8~{\AA} (5.7~meV).  The spectrometer was in low-resolution mode with a 1/2 frame overlap ratio and a minimum sample-to-detector time of 1200~$\mu$s.  All samples were measured in closed-cycle refrigerators, with the field dependence measured in a 10-T superconducting magnet.

\begin{figure}
\centering
\includegraphics[width=7cm] {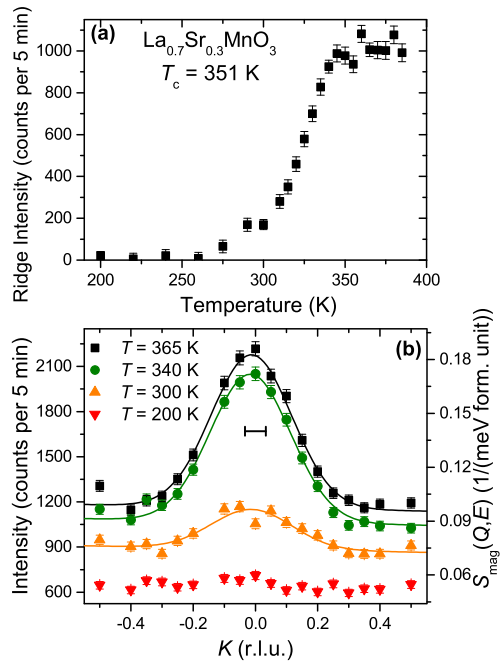} \vspace{-4mm}
\caption{(a) Temperature dependence of the ridge intensity in $x$=0.3~LSMO, measured at (1.4~0~0) with $E$~=~3~meV.  (b) Transverse (1.4~$K$~0) scans at $E$~=~3~meV.  The lines are fits to Gaussian line shapes with a HWHM of 0.152 r.l.u. plus a sloped background.  The instrumental resolution (FWHM) is shown as the horizontal black bar.  Error bars and uncertainty values throughout this manuscript are statistical in nature and represent one standard deviation.}
\label{LSMO30}
\end{figure}
The temperature dependence of the inelastic ridge intensity at $E$=3~meV is displayed in Fig.~\ref{LSMO30}(a) for $x$=0.3~LSMO.   The ridge position is at (1.4~0~0), and we have subtracted a background that was determined by averaging scattering at (1.4~$\pm$0.5~0). This magnetic scattering ridge first appears at a temperature of around 275~K (0.78~$T_c$) and its intensity increases steadily with increasing temperature until $T_{c}$ is reached; the ridge intensity remains constant with increasing temperature for temperatures above $T_{c}$ up to at least 385~K.  Figure~\ref{LSMO30}(b) shows transverse scans through the ridge position.  The ridge intensity has been fit to a Gaussian line shape with a HWHM of 0.152~reciprocal lattice units (r.l.u) or 0.246~{\AA}$^{-1}$; there is no discernable temperature dependence to this width.  Similar measurements on $x$=0.3 LCMO and LBMO as well as $x$=0.2~LCMO all display qualitatively similar behavior.  In each of these samples the magnetic ridge has a width of 0.152~r.l.u. and an intensity that increases with increasing temperature from around ($T_c-$75~K) to $T_c$ and is temperature independent above $T_c$.

This temperature dependence is similar to that of diffuse polaron scattering \cite{Adams2000,Lynn2001}, measured, for example, at the (1.85~2~0) position near a fundamental Bragg peak. This structural (lattice) intensity is associated with scattering from a single polaron and also was relatively temperature independent above $T_c$.  Diffuse scattering from uncorrelated polaron distortions is a common feature of CMR compounds, including the nearly half-doped (Nd$_{0.125}$Sm$_{0.875}$)$_{0.52}$Sr$_{0.48}$MnO$_3$ \cite{Shimomura} and the bilayer manganite La$_{1.2}$Sr$_{1.8}$Mn$_2$O$_7$ \cite{VasiliuDoloc1999}.  However, the diffuse structural scattering appears only in a narrow temperature range just below $T_c$, while the magnetic ridge scattering in all of these materials develops over a wider range of about 75~K.  This suggests that the magnetic part of the polarons - the conduction electron hopping between ionic sites - first arises at lower temperatures while the lattice part of the polaron (the Jahn-Teller distortion surrounding the conduction electron) becomes coupled or develops enough strength to be noticeable in a neutron scattering experiment only as the temperature approaches $T_c$. Another possible explanation for this temperature dependence is that the magnetic ridge intensity below $T_c$ could represent polarons self-trapped in a paramagnetic volume fraction embedded into a phase-separated ferromagnetic primary phase \cite{Uehara}.

\begin{figure}
\centering
\includegraphics[width=7cm] {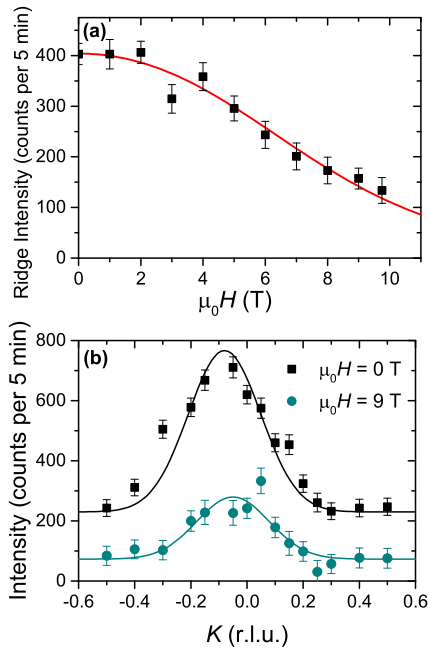} \vspace{-4mm}
\caption{(a) Magnetic field dependence of the ridge intensity in $x$=0.3~LCMO, measured at (1.4~0~0) with $T$~=~270~K and $E$~=~2~meV.  The field was applied along [001].   The line is a fit as described in the text. (b) Transverse (1.4~$K$~0) scans both at zero field and $\mu_{0}H$=9~T.  The lines are Gaussian fits with a HWHM of 0.152 r.l.u.}
\label{Field}
\end{figure}
Figure~\ref{Field}(a) displays the magnetic field dependence of the ridge intensity in $x$=0.3~LCMO, measured at 270~K with $E$=2~meV.  The ridge position is at (1.4~0~0), and we have subtracted a background that was determined by averaging scattering at (1.4~$\pm$0.5~0).  The ridge intensity is decreased due to the applied field, as would be expected for a scattering component associated with the paramagnetic phase.  However, this effect is fairly weak with about one-third of the ridge intensity remaining even under an applied field of 9.75~T.  To confirm that the ridge intensity is still present under high applied magnetic fields, Fig.~\ref{Field}(b) displays transverse scans through the ridge position both at zero applied field and with $\mu_{0}H$=9~T.  For both of these scans, an additional background was subtracted which was determined by a scan at 125~K in zero field.  Previous reports have revealed the magnetic field dependence of both polaron correlations \cite{Lynn2001} and the quasielastic central component \cite{Lynn1996,Lynn1997} of $x$=0.3~LCMO at temperatures near $T_c$.  An applied field of 6~T is enough to suppress the scattering in those cases; even smaller applied fields will suppress the diffuse polaron scattering in bilayer manganites such as La$_{1.2}$Sr$_{1.8}$Mn$_2$O$_7$ \cite{VasiliuDoloc1999}.  The magnetic field strength required to suppress these ridges of scattering is also large compared to the fields associated with colossal magnetoresistance in these materials \cite{Schiffer}.  These data suggest that the magnetic fields needed to significantly alter the number of polarons present are significantly larger than the fields needed to disrupt correlations between polarons.  This broad field dependence is consistent with the 75~K transition range in the temperature dependence of the magnetic ridges and with the bulk magnetization of $x$=0.3~LCMO, which at $T$~=~270~K and $\mu_{0}H$=9~T is approximately 70\% of the low-temperature saturated magnetization.

\begin{figure}
\centering
\includegraphics[width=7cm] {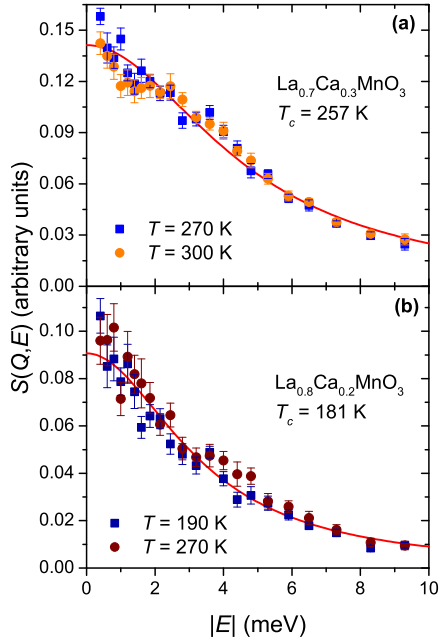} \vspace{-4mm}
\caption{(a) Energy dependence of the scattering ridge in La$_{0.7}$Ca$_{0.3}$MnO$_{3}$ at 270~K and 300~K. The line is a fit to a Lorentzian line shape with $\Gamma$=4.7$\pm$0.1~meV.(b) Energy dependence of the scattering ridge in La$_{0.8}$Ca$_{0.2}$MnO$_{3}$ at 190~K and 270~K.  The line is a fit to a Lorentzian line shape with $\Gamma$=3.3$\pm$0.1~meV.  Using data measured on the DCS spectrometer, transverse cuts through the ridges were made for the displayed energy transfers and with $q$ integrated over $\pm$0.24~{\AA}$^{-1}$ along the ridge direction.  These cuts were fit to a Gaussian lineshape, and the integrated area of the ridges is shown as a function of the energy transfer.}
\label{EnergyDep}
\end{figure}

The energy dependence of this quasielastic scattering is shown, for $x$=0.2 and $x$=0.3 LCMO, in Fig.~\ref{EnergyDep}.  The data have been fit to Lorentzian lineshapes with a HWHM of $\Gamma$=4.7$\pm$0.1~meV for the $x$=0.3 sample and $\Gamma$=3.3$\pm$0.1~meV for the $x$=0.2 sample.  There is no significant change in the energy dependence as the temperature is increased from slightly above $T_c$ to as high as 1.5$T_c$ in the $x$=0.2~sample.  $\hbar/\Gamma\approx$200~fs can be considered as roughly the time scale over which these correlations persist.

\section{Discussion and Conclusions}

\begin{figure}
\centering
\includegraphics[width=8cm] {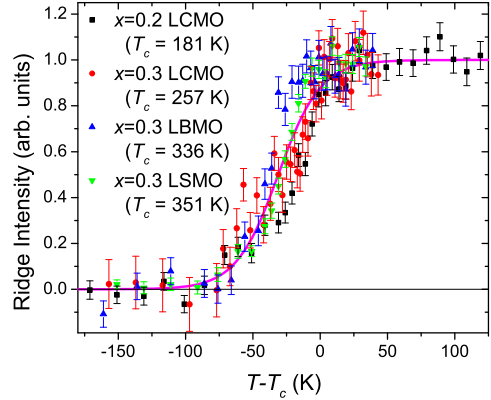} \vspace{-4mm}
\caption{Normalized ridge intensity as a function $T-T_c$ for four samples.  The line is a guide to the eye.}
\label{TempCompare}
\end{figure}

The ground states of $x$=0.2 LCMO (ferromagnetic insulating) and the three $x$=0.3 samples (ferromagnetic metallic) differ substantially, yet the temperature dependence of the magnetic ridge intensity is qualitatively quite similar across all of these samples.  Figure~\ref{TempCompare} compares these samples by plotting the normalized ridge intensity against $T-T_c$.  This shows that, in all four samples, the magnetic ridge intensity develops over a temperature range of about 75~K. Both metallic and insulating samples show evidence of polaron behavior near $T_c$ \cite{Dai2000,Dai2001} and the short-range ferromagnetic correlations of self-trapped magnetic polarons should diminish in both types of samples at temperatures well below $T_c$ but by differing mechanisms.  In CMR materials such as the $x$=0.3 samples, the short-range hopping character of the conduction will become less pronounced at lower temperatures as long-range ferromagnetic order allows for metallic conduction.  In $x$=0.2 LCMO the phase transition instead yields an insulating ferromagnetic ground state, driven by the suppression of coherent Jahn-Teller distortions induced by orbital ordering \cite{VanAken}.  The short-range magnetic order arising from polaron hopping is then suppressed as orbital order inhibits the cooperative lattice distortions necessary for colossal magnetoresistance \cite{Weber}.

This 75~K temperature range over which the magnetic ridges develop is large in comparison to the sharp temperature range over which both the quasielastic central component and diffuse polaron scattering develop in $x$=0.3 LCMO \cite{Adams2000}, but is still smaller than the 100~K to 180~K range of temperatures below $T_c$ for which $x$=0.3 LBMO and LSMO display dynamic polaron correlations in the ordered state \cite{Chen}.  The fact that this scattering component develops over a temperature range of about 75~K in all of these samples means that in the highest-$T_c$ sample ($x=0.3$ LSMO) the magnetic ridge intensity first appears upon warming to about 0.8$T_c$ while in the lowest-$T_c$ sample ($x$=0.2 LCMO) the ridge intensity appears at temperatures as low as 0.55$T_c$.  If the magnetic superexchange constant, $J$, were the only relevant energy scale, then ridge intensity would likely display universal behavior as a function of the reduced temperature ($T/T_c$).  This is not the case, consistent with results that polaron formation influences CMR manganites in a way that breaks conventional relationships among $J$, $T_c$, and the spin wave stiffness coefficient \cite{Adams2004,FernandezBaca,Zhang}.

A simple model of magnetic polaron hopping can also describe the magnetic field dependence of the ridge scattering displayed in Fig.~\ref{Field}(a).  The resistivity of CMR manganites has been analyzed \cite{Wagner} by assuming that carrier hopping is thermally activated with an energy barrier of $W$ and that this energy barrier is lowered when neighboring spins become ferromagnetically aligned as
\begin{equation*}
W \rightarrow W_0\left[1-\textrm{tanh}^2\left(\frac{g\mu_BHS}{k_B(T-T_c)}\right)\right]
\end{equation*}
where $\textrm{tanh}\left(\frac{g\mu_BHS}{k_B(T-T_c)}\right)$ is a modified Brillouin function to describe the relative magnetization of $S$=3/2 spins in a ferromagnetic material above $T_c$.  As the magnetic field is increased, the hopping barrier approaches zero as the material moves toward the metallic ferromagnetic phase.  If we assume that the magnetic ridges arise from magnetic polarons that become self-trapped after a small number of hops, then the ridge intensity should decrease as the material becomes ferromagnetically aligned with a field dependence that can be modeled as
\begin{equation*}
I(H)\propto1-\mathrm{exp}\left(-\frac{W_0}{k_BT}\left[1-\textrm{tanh}^2\left(\frac{g\mu_BHS}{k_B(T-T_c)}\right)\right]\right).
\end{equation*}
The data in Fig.~\ref{Field}(a) have been fit to this equation with the only fit parameter (other than a scale factor) fitting to $W_0=32.9\pm5.8$~meV.  This behavior is reminiscent of the universal relationship between magnetization and lattice distortions observed in La$_{1-x}$Ca$_x$MnO$_3$ \cite{Bridges}.

A scan through the elastic peak of a 0.0414~mol vanadium standard was performed immediately after, and in the same spectrometer configuration as, the measurements on the $x$=0.3 LSMO sample.  The incoherent cross section of vanadium was used to normalize the measured count rate to determine $S_{mag}(Q,E)$ in absolute units of $(\mathrm{meV}\cdot\mathrm{formula~unit})^{-1}$; this absolute scale is displayed on the right-hand axis of Fig.~\ref{LSMO30}(b).  Using this absolute scale the intensity of the magnetic ridge can be integrated over all energy transfers (assuming a Lorentzian line shape with $\Gamma$=3.3~meV) and over all of momentum space within one Brillouin zone (assuming that each of the six ridge directions extends a longitudinal distance of 0.20~{\AA}$^{-1}$ from the edge of the ring before reaching the zone boundary).  This is compared to the sum rule
\begin{equation*}
\int_{-\infty}^{\infty}dE \int_{BZ}dQ \, S_{mag}(Q, E) \, = \, \frac{2}{3} \frac{(2\pi)^{3}}{v_{0}}S(S+1)
\end{equation*}
where the factor of $\frac{2}{3}$ arises because neutron scattering is sensitive only to the spin components perpendicular to $\vec{Q}$.  We find that the magnetic spin associated with this scattering is given by $S(S+1)$=0.45$\pm$0.03.  The total scattering associated with this magnetic ridge is then about 8.5\% of the total magnetic scattering expected for the system.

\begin{figure}
\centering
\includegraphics[width=7cm] {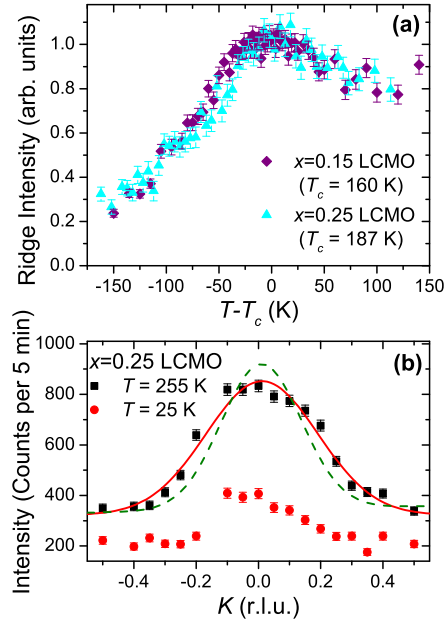} \vspace{-4mm}
\caption{(a) The normalized ridge intensity as a function $T-T_c$ for both $x$=0.15 and $x$=0.25 LCMO.  (b) Transverse (1.4~$K$~0) scans at $E$~=~3~meV for $x$=0.25 LCMO at $T$ = 255~K and 25~K.  The lines are fits of the 255~K data to a Gaussian plus a sloped background.  For the solid red line, the HWHM of the Gaussian is 0.212 r.l.u.  For the dotted green line, which is a poor fit to the data, the Gaussian has a HWHM of 0.152 r.l.u. identical to that observed the the optimally doped samples.}
\label{Inhomogenous}
\end{figure}

The temperature dependence of the magnetic ridge for both metallic $x$=0.25 LCMO ($T_c$=187~K) and insulating $x$=0.15 LCMO ($T_c$=160~K) is displayed in Fig.~\ref{Inhomogenous}(a); the data were measured at an energy transfer of $E$=3~meV and the ridge intensity is defined as the difference in scattering intensity at (1.4~0~0) and the average background intensity at (1.4~$\pm$0.5~0).  These data show several qualitative differences from those measured on the near-optimally doped samples or on the $x$=0.2 LCMO sample.  First, measurements at very low temperatures show that a weak ridge remains present far below $T_c$.  The temperature dependence of this ridge intensity, in sharp contrast to the previous samples, increases steadily from a nonzero value at $T\approx0$ to a broad maximum at $T\approx T_c$ and decreases slowly as the temperature is further increased.  Second, the magnetic ridges observed in these samples are broader in the transverse direction than those observed in other samples.  In particular the transverse scan in the $x$=0.25 sample, displayed in Fig.~\ref{Inhomogenous}(b), is clearly inconsistent with the 0.152~r.l.u. HWHM that had been observed in the optimally doped samples.  Because this behavior is observed in the $x$=0.15 and $x$=0.25 LCMO samples but not the $x$=0.2 sample these differences may arise from other effects, such as sample homogeneity, rather than the doping level.

Intrinsic chemical inhomogeneity is known to play an important role in La$_{1-x}A_x$MnO$_3$ samples with $x\lesssim$0.3, with nanoscale clustering of $A$ dopants \cite{Shibata,Rozenberg2007,Rozenberg2011}.  This nonrandom distribution is more pronounced in samples with smaller doping levels and much more pronounced in Ca-doped samples than in Ba-doped samples \cite{Auslender}.  It is possible that the $x$=0.15 and $x$=0.25 LCMO samples reported here display more nanoscale nonrandom chemical clustering than the other samples.  This would be consistent with the observation that the $x$=0.25 LCMO sample measured here displayed a significantly lower $T_c$ than powder samples or other single crystal samples of the same nominal concentration \cite{Huang,Dai2001}, while the $x$=0.2 and $x$=0.3 LCMO samples measured here displayed a comparable or higher $T_c$ than other reported single-crystal samples of the same nominal concentration \cite{Okuda,Dai2001}. While chemical inhomogeneity might result in a spread of transition temperatures, that alone is unlikely to explain the temperature dependence of the magnetic ridge intensity in these samples.  Inhomogeneity in manganites has been shown to reduce the energy barrier for polaron formation even when the exchange interaction is only slightly lowered \cite{Sato}.  Local lattice distortions due to non-random $A$-site clustering could trap polarons at lower temperatures; a range of local distortions could yield the very broad temperature dependence observed in $x$=0.15 and $x$=0.25 LCMO.  Similarly, nanoscale chemical inhomogeneity might disrupt the longer length-scale phase coexistence caused by quenched disorder that is believed to play a role in the CMR effect \cite{Burgy,Moreo,Uehara}.  Nonrandom distribution could explain the unusually large width observed for transverse scans through the magnetic ridge in these samples, as significant clustering of the Ca dopants might restrict the ferromagnetic hopping to a smaller spatial region.

In conclusion, ridges of quasielastic magnetic scattering are observed in La$_{1-x}A_{x}$MnO$_{3}$ materials displaying both metallic and insulating ferromagnetic ground states.  This scattering component is roughly temperature independent above $T_c$, similar to the temperature dependence of diffuse (uncorrelated) polaron scattering in these materials.  A simple model of short-range ferromagnetic correlations dependent upon the number of nearest-neighbor hops connecting two Mn ions reproduces the essential features of this magnetic scattering.  We therefore associate the ridges of magnetic scattering with the magnetic part of diffuse polaron scattering resulting from short-range ferromagnetic correlations caused by a conduction electron hopping between Mn ions.  This magnetic component of the nanoscale polarons, common to ferromagnetic manganites, complements the previously observed lattice polaron components to further reveal the coupling between the lattice and magnetic degrees of freedom in colossal magnetoresistive materials and other ferromagnetic manganite compounds.

\section*{ACKNOWLEDGEMENTS}
This work utilized facilities supported in part by the National Science Foundation under Agreement No. DMR-0944772.  J.S.H acknowledges partial support from the Office of Naval Research through the Naval Academy Research Council, award $\#$N001614WX30023. D.A.S. acknowledges in part the Ministry of Education and Science of the Russian Federation, Project No. 3.2076.2014/K, designated within the framework of 2014-2016 competitive grants for universities.\\ \\
\\
$\ast$ Email address: helton@usna.edu\\
$\ddag$ Email address: jeffrey.lynn@nist.gov\\
$\dag$ Deceased

\bibliography{LCMO_PRB}
\end{document}